# Iron phosphate glass structure at different length scale with emphasis on the medium range: a classical molecular dynamic study

Shakti Singh[a], Manan Dholakia[b] and Sharat Chandra[b]

[a] Photonics Materials Technology Section, Materials Science and Advanced Technology Group, Raja Ramanna Centre for Advanced Technology, a CI of Homi Bhabha National Institute (Mumbai), Indore, MP, 452013, India

[b] Materials Science Group, Indira Gandhi Centre for Atomic Research, a CI of Homi Bhabha National Institute (Mumbai), Kalpakkam, TN, 603102 India

E-mail: corresponding author S.S: shaktis@rrcat.gov.in,

M.D: manan@igcar.gov.in, S.C: sharat@igcar.gov.in

## Abstract

Glasses are known to have medium range order (MRO) but its link to any experimentally measurable quantity is still ambiguous. The first sharp diffraction peak (FSDP) in structure-factor $S(q)$ obtained from diffraction experiments on glasses has been associated with this MRO (~ 7-15 Å) but understanding the fundamental origin of this universal peak is still an open problem. We have addressed this issue for a complex glass i.e. iron phosphate glass (IPG) through atomistic models generated from a hybrid approach (our in-house developed MC code + MD simulation). IPG is a technologically important glass with applications in waste vitrification, bioactive glass, laser glass material, anode material for batteries etc and is seen as a strengthened substitute for borosilicate glasses. We have performed a comparative study by generating glass models from different initial configurations and randomization techniques. The developed IPG models were first validated with existing data on short range order (SRO) and MRO, through study of pair correlation functions, bond angle distributions and coordination number (CN) for SRO and rings distribution, FSDP in structure factor and void size distribution for MRO. The study of coordination environment of oxygen is specifically shown to aide in understanding glass formation through topological constraint theory. Thereafter to understand the fundamental origin of FSDP in $S(q)$, structure factors were calculated corresponding to individual ring sizes present in the model. The relative contribution of these individual $S(q)$'s in the total experimental $S(q)$ is estimated using an inverse fitting approach. The contributions thus obtained directly correlated with rings size percentages in the models for the considered q-range. In particular, the melt-quenched model obtained from MC model as initial structure is found to reproduce most experimental features seen in IPG. Through this exercise we can connect the rings distribution of an atomistic glass model with an experimentally measurable quantity like FSDP in S(q) for a complex glass like IPG. This gives physical meaning to the rings distribution while also proving that this structural descriptor is a useful tool for validation of MRO in simulation-produced models of glass.

# I. Introduction

Glasses are versatile. They have various applications in the fields of optical communication, spectroscopic instruments, packaging, smartphones, radioactive waste vitrification, battery materials, biomedical applications and many more. Chemical inertness, hardness and the ability to accommodate stress for long times are the stand-out properties of glasses that enable their vast usage. Yet on the atomic scale its structure is the subject of century-old research [1]. The past few decades have witnessed resurgence in research on this subject because of the advancements in simulations and computational power. The atomistic models that these simulations produce, allow for physical insights on the observed properties and also predictively explore the properties of glasses (chapter 21 of [2]). Due to the random nature of glass structure, almost all experimental probes fail to successfully capture the medium-range structure, which is why the research in this area is mostly from simulations. So, two important aspects of glass modelling are the nature of medium range order (MRO) and the effect of various simulation routines on the generated glass models. In this study, the former is addressed via state-of-the-art tools like rings analysis, voids analysis, FSDP analysis, while for the latter we have prepared glass models using different initial configurations and randomization techniques. This study presents classical molecular dynamical based modelling of iron phosphate glass (IPG). Simulation studies on IPG also has a rich history wherein both classical molecular dynamical ( [3–8] ) and ab-initio DFT ( [9–12] ) based methods are used to gain insights on IPG structure. In this study, different initial configurations are employed which can be categorized as crystalline, randomly-decorated and totally randomized. While the crystalline configuration mandatorily requires melt-quench simulation for glass generation, the randomly-decorated configuration can either be just equilibrated or melt-quenched to generate glass. So, both these variations are studied. The third category is taken from the literature [8] wherein starting from a totally randomised atomic configuration glass models are generated using melt-quench simulations.

The FSDP in structure factor of glasses has been linked to MRO which has various interpretations: it is shown to be arising from chemical correlations between atoms [13]. The linkages of cavities/void space left due to inefficient random packing in glasses [14–16] has also been linked to FSDP. The void/cavity-based explanation, though intuitive can only be studied using positron annihilation lifetime spectroscopy, which places limitations on the size of the voids that can be characterised. On the other hand, rings or chain like repeating clusters are also proposed to explain the MRO [17,18]. Among the various interpretations, rings based explanation has recently been validated against experimental data [17]. The existence of this topological feature in disordered solids was proposed theoretically by Zachariasen [19] in 1920s and then experimentally observed only in 2012 from direct imaging of two-dimensional silica using atomic resolution transmission electronic microscopy [20]. Recent studies [17,21,22] on silica glasses have also reported a similar correlation between theoretically calculated rings distribution and the FSDP (in particular) in S(q). The present study attempts a similar correlation on much more complex glass, IPG, over an extended q range. The rings characterization of a disordered solid is a state-of the-art tool to describe the MR structure and has advanced from a purely theoretical and simulation-based concept to experimentally verifiable and physical property linked concept. This claim is supported by some recent studies [23–27] that highlight the rings structure-property correlation in disordered solids. Further proof of this concept is available in 2D glasses which can be accurately imaged leading to prediction of the correlation between rings topology and physical properties [28–30]. The boom in number of studies in this topic came after the continuous random network (CRN) theory of Zachariasen was confirmed experimentally using 2D silica glass, more than 80 years after its proposal [20].

Hence, it is highly desired that a link between the rings size distribution (theoretical construct) in an atomistic model and an experimental observable should be made to better validate the theoretical models. In the present study, the individual $S_i(q)$'s are used to fit a non-linear combination of these to experimental S(q) and subsequently their fitted-weighting coefficients are compared with rings size distribution data calculated from simulation models. In this way, the experimental structure factor data is used to validate the MRO present in our models through calculation of the theoretical construct i.e., the rings distribution.

Since ab-initio based methods like DFT have computational bottlenecks on the size of atomistic models produced, the present study only discusses larger models developed using MD. This requirement is a must since the MR structure is the focus of this study and previous studies on silica-based glass suggest the model size should be minimum 20 Å cube for it to correctly produce the MRO [13,31]. Although a size-dependent study for IPG may be warranted for arriving at such a limit for IPG, this is not the scope of this study.

The paper is organized as follows: the next section on computational procedures summarizes the techniques adopted for modelling and various models considered in this study. This is followed by results and discussion section which first presents the short-range (SR) properties i.e. pair distribution functions, bond-angle distributions and coordination number (CN), geometry and environment around oxygen. Subsequently the important MR properties like structure factor, FSDP and rings topology are discussed. Finally, the conclusion summarizes the findings of this study.

## II.  Computational procedure

The IPG models developed in this study follow a 40:60 (by mole %) $Fe_2O_3:P_2O_5$ composition where the Fe/P ratio is 0.75. A similar composition crystal with formula $Fe_3(P_2O_7)_2$ [32], in which the Fe/P ratio is 0.75 is used for finding similarity with rest of the glass models developed in this study. This is labelled as "$Fe_3(P_2O_7)_2$ crystal" and has dimensions 9.08 Å x 10.33 Å x 12.40 Å of the orthorhombic unit cell. The structural model is obtained from materials project website [33] (*Data retrieved from the Materials Project for $Fe_3(P_2O_7)_2$ (mp-559713) from database version v2023.11.1*). Also, starting with a 2×2×2 supercell of this crystal, melt-quench simulation is performed to obtain glass model for IPG using MD, labelled as "MQ_crystal" in this study. The starting model generated using Monte-Carlo method [34], is labelled "MC_as_dev.", which is a cube of edge 20 Å comprising 411 O, 114 P and 87 Fe atoms. Then a 2×2×2 simulation cell of the "MC_as_dev." model is equilibrated at 300 K for 60 ps using MD and is called "MC_eq_MD", while the model obtained by again performing the melt-quenching MD procedure starting from the equilibrated simulation cell is called "MC-MQ". A very large model (cubic cell with edge of 90 Å) built using MD by melt-quenching technique developed by Pawel Stoch et al. (labelled as "MQ_Stoch et al." model), the only other MD model available in the literature, is also used for comparison throughout the study. Apart from this, experimental data wherever available is also used to validate the simulation-based results obtained in this study.

The initial random structures obtained from MC were then melt quenched, using molecular dynamics (MD) simulations as implemented in LAMMPS [35]. The interatomic potential for IPG was first developed by Bushra et al. [3] and later modified by Kitheri et al. [4]. This potential consists of two body interactions, modelled using a Buckingham potential [36], together with a Coulomb term to model the long-range interactions between ionic charges.

$$V(r_{ij}) = \frac{q_i q_j}{4\pi\epsilon_0 r_{ij}} + A_{ij} exp\left[-\frac{r_{ij}}{\rho_{ij}}\right] - \frac{C_{ij}}{r_{ij}^6} \quad , \quad r > 1 \text{ Å} \quad (1)$$

Here, $A_{ij}$, $\rho_{ij}$ and $C_{ij}$ are constants and $r_{ij}$ is the distance between pair of atoms $i, j$. First term is the Coulomb term, the second term is the repulsive term arising due to the interpenetration of the electron shells of the atoms at small enough distances and the third term is the attractive term similar to the Lennard-Jones potential [37]. The exponential term in equation (1) converges to a constant, while the attractive term blows up as $r_{ij} \to 0$. Hence this potential becomes attractive at very small distances. This problem is obliterated if we use the ZBL potential [38] for describing the interactions at $r_{ij} < 1$Å. The ZBL potential is a screened Coulomb function given by,

$$V(r) = \frac{1}{4\pi\epsilon_0}\left(\frac{Z_i Z_j e^2}{r}\right)\psi\left(\frac{r}{a}\right) \quad (2)$$

where $Z_i$ and $Z_j$ are the nuclear charges on ions $i$ and $j$, $\psi$ is the screening function, $a$ is the screening parameter and other things have their usual meaning. The three body interactions are modelled using Stillinger Weber potential [39] given by,

$$V_3(r_{ij}, r_{ik}, \theta_{jik}) = \lambda \, exp\left(\frac{\gamma}{r_{ij} - r_c} + \frac{\gamma}{r_{ik} - r_c}\right)(cos\theta_{jik} - cos\theta_0)^2 \quad , \quad r > 1 \text{ Å} \quad (3)$$

where for the atom triplet $j - i - k$, $r_{ij}$ and $r_{ik}$ are the two internal atomic separations, and $\theta_{jik}$ is the bond angle formed at the central atom $i$. $r_c$ is the cut-off radius beyond which the three body terms are set to zero. $\lambda, \gamma$ and $\theta_0$ are empirical parameters. The three body potential in IPG is used to control the local bond angles in the triplets O-P-O and P-O-P. Hence it ensures the stability of PO$_4$ tetrahedra and allows it to form glassy chains within the IPG network.

A 2×2×2 supercell of the MC_as_dev. model corresponding to size $40 \times 40 \times 40$ (Å$^3$) was taken, and to integrate the equations of motion the velocity-verlet [40–42] algorithm, with a timestep of 0.1 fs was used. To model the process of melt quenching the methodology similar to Deng et. al [43] was used. The system was first relaxed at 0 K using Conjugate Gradient method [44]. Then, NVT ensemble was used to first equilibrate it at 300 K for 60 ps, then heated to 6000 K for the next 60 ps. The temperature of the system was subsequently brought down to 5000 K, followed by equilibration for 100 ps. The completely melted system was then cooled down at a quenching rate of 5 K/sec, till the temperature became 300 K. For the next 20 ps, the system was equilibrated at 300 K using NVT, and followed by NVE ensemble for the last 10 ps. Thus, highly randomized, amorphous structures of IPG were obtained.

# III. Results and discussion
## A. Short-range properties

The structural properties are divided based on the length scale they characterize i.e. short range (SR) for 0-5 Å, medium range (MR) for 5-15 Å and long-range (LR) for >15 Å. We have first presented the short-range properties of the models. Figure 1 shows the total pair correlation function (PCF) of IPG models considered in this study. The development of peaks at 1.5 Å, 1.8 Å, 2.5 Å, 3.25 Å and 4 Å can be seen. The prominent peaks are the first three which correspond to P-O, Fe-O and O-O distance correlations. The P-O and Fe-O peaks arise due to existing covalent and ionic bond between these species. The Fe-O peak for crystal $Fe_3(P_2O_7)_2$ is split into two peaks corresponding to two different bond distances of Fe-O i.e. Fe-O bond distance of 1.9 Å for exclusive P-O-Fe type connections (only corner sharing O i.e. CN(O)=2) and 2.2 Å for P-O-$Fe_2$ type connections (i.e. CN(O)=3) that share a common face between two Fe polyhedral units. Refer to Figure 9 (a) for visualisation. Although discriminated using the crystal system, these connection types are present in glass models as well but are rather distributed in the range (1.75 Å, 2.3 Å) resulting in a single broad peak around 1.9 Å. These two different bond-lengths have implications on the glass strength as well, wherein the P-O-Fe type of connections make the IPG more chemically durable, an excess of them results in P-O-$Fe_2$ type of connections which overburdens the network resulting in weak strength of the network, as also represented by increased bond-length of this type of connections. Next, the O-O peak at 2.5 Å arises due to the presence of O at vertices of the regular coordination polyhedra that the P and Fe cations form throughout the network. More about this peak will be discussed in the partial pair correlation function (PPCF) discussion corresponding to it. Thereafter the diminished peaks at 3.25 Å and 4 Å arise from second neighbour correlations due to the repeating connections of a particular type. For eg., the P-P correlation arises from $P_2O_7$ type dimeric connection between two phosphorous atoms, that gives a peak at 3 Å. Similarly, the presence of Fe polyhedra at many of the vertices of the P-O tetrahedra results in the Fe-P second neighbour correlation peak observed at 3.25 Å. Similarly, Fe-Fe correlation results in a broad peak at 3-4 Å, more about which will be discussed while reporting the bond angles.

A general observation can be made that prominent peaks in the PCF of IPG develop around the peaks present in the crystalline system as well, although less intense and broader. The PCF of all the glass models averages to 1 after ~4.5 Å showing the presence of disorder leading to averaging of any subsequent distance correlations among atoms, whereas the crystalline systems give very prominent peak even beyond this length scale.

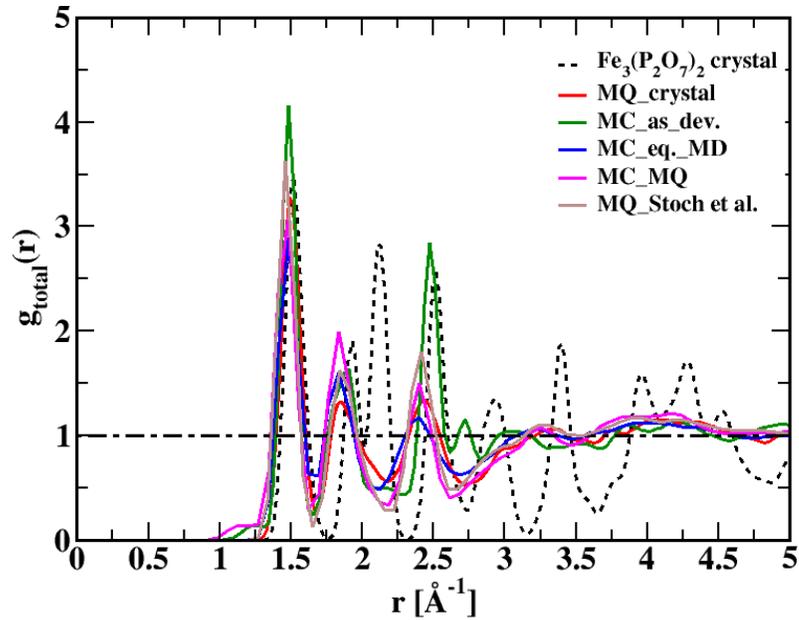

*Figure 1: Total pair correlation function of IPG models considered in this study*

Next, in Figure 2 the PPCF curves are presented. The curves can be better appreciated when seen with the nearest and next-nearest neighbour distances shown in Figure 3. In Figure 2(a), the PPCF g(r)_P-O is shown. The peak at 1.51 Å is exactly like that seen in crystal as well, showing that the P atoms maintain the sanctity of their covalently bonded tetrahedra even in glass models. Also, there is no asymmetry in this peak, implying that P=O i.e terminal oxygen are absent at this composition. The peak bond-length value has excellent agreement with the experimental value of 1.53 ± 0.01 Å found using neutron diffraction (ND) data [45]. In Figure 2(b), the g(r)_Fe-O is shown. The reason behind two distinct peaks in crystal system is already discussed earlier which leads to a broad first peak in the glass models at 1.9 Å, which compares very well with experimental data from literature (ND 1.94±0.02 Å [45] and from EXAFS 1.89±0.01 Å [46]). In Figure 2(c), the first O-O peak also features a diminished secondary peak near its tail. As can be seen in Figure 3, the main peak is due to 3(i) type O-O distance from P tetrahedra whereas the secondary peak is from 3(ii) type O-O distance from Fe polyhedra. The former peak occurs at 2.4 Å which agrees well with experimental value of 2.52±0.02 Å [45]. The broadness of the secondary peak also represents that there is significant distortion in the Fe polyhedra (which comprises a mix of distorted tetra, penta and hexa-coordinations) as compared to the crystal system where only hexa-coordination is present (octahedral and trigonal prismatic geometry, see Figure 9). Next the P-P peak in Figure 2(d), as discussed earlier, arises from P-P correlation in $P_2O_7$ dimeric units, shown as type 4 in Figure 3. In Figure 2(e), the g(r)_Fe-Fe shows a very faint peak at ~3.25 Å. The peak in crystalline configuration is due to Fe-Fe correlation between the face shared octahedral and bipyramidal units. This correlation is difficult to observe in the glass models as higher order correlations tend to average out in glasses. Nevertheless, the faint correlation present is due to the corner sharing Fe polyhedra shown in Figure 3 and also the edge sharing polyhedra shown in Figure 5, later. The peak is found at ~3.25 Å for MC_eq._MD model, MC_MQ and the MQ_Stoch et al. model which compares well with EXAFS obtained value ( 3.67±0.01 Å [46]) Lastly the Fe-P correlation resulting in peak in g(r)_Fe-P in Figure 2(f) at 3.25 Å can also be seen in δ-type distances in Figure 3. This value also compares well with the experimental results (3.25 Å [45]).

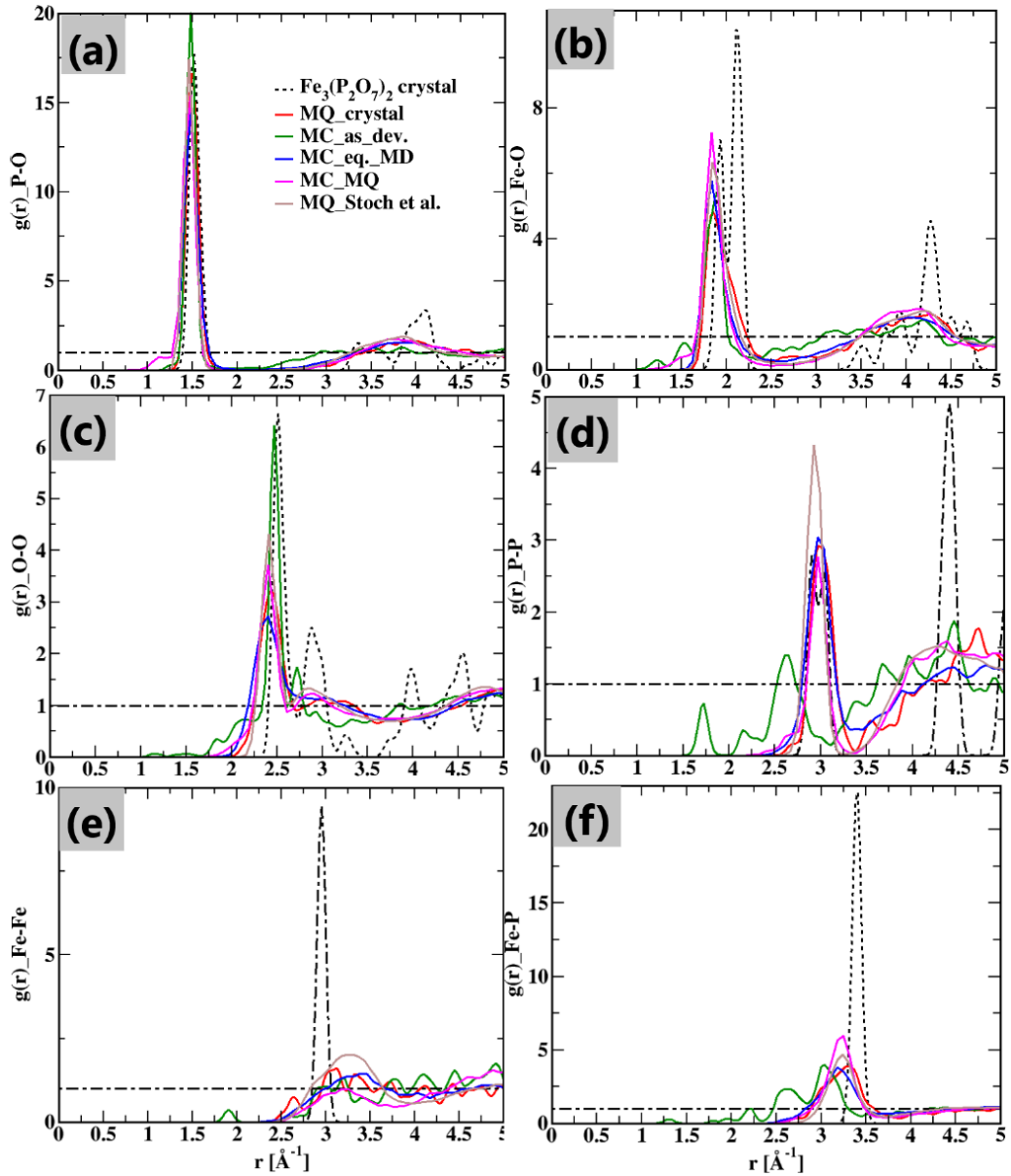

*Figure 2: Partial pair correlation function of IPG models considered in this study*

From the PCF and PPCF just presented, important observations can be made regarding the role of initial structures in glass modelling when the short-range structural properties are required to be reproduced. In this study, the starting structures used were the $Fe_3(P_2O_7)_2$ crystal and MC_as_dev. model. The two starting models produced very much the same PCF and PPCF. The only difference that can be observed is in the next nearest neighbour pair correlations like P-P and Fe-Fe, where the MQ_crystal model shows ripples whereas the MC_MQ model is smoother. The difference can be due to preservation of some crystalline signatures at higher correlation lengths even after melt quenching. Also, at this level of order, it is very hard to distinguish between the equilibrated MC model and MC_MQ model. So just equilibration of the MC model produces PCF and PPCF curves that are in very good agreement to MC_MQ model.

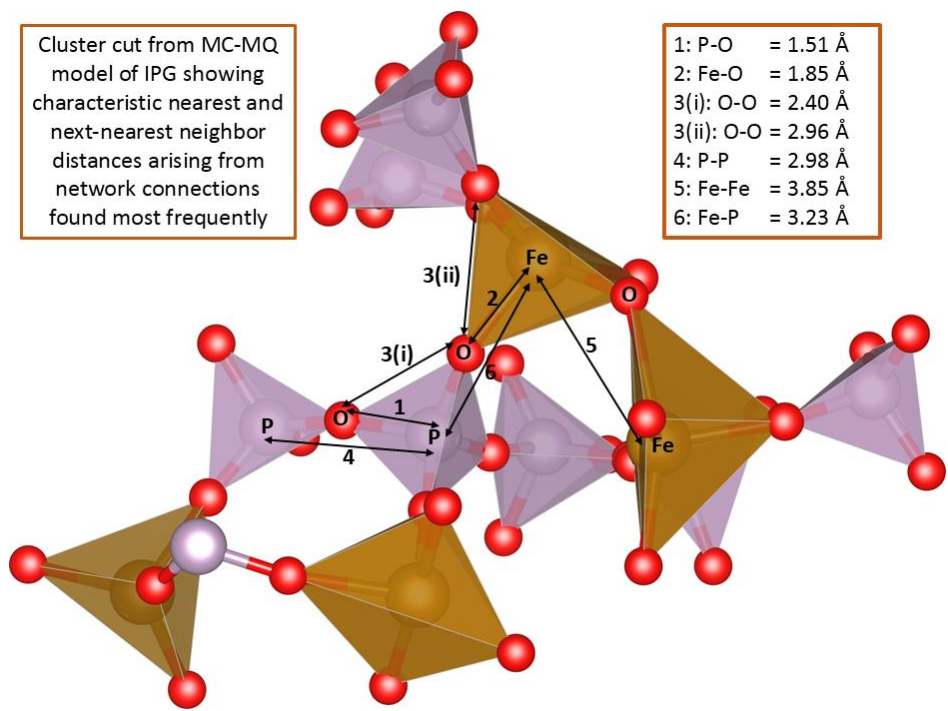

*Figure 3: Cluster of atoms in IPG showing characteristic distances between Fe, P and O atoms. The distance values shown in the box are for this snapshot. Visualization by VESTA [47]*

Next, the bond angle distributions are shown in Figure 4. In Figure 4(a), the O-P-O bond angle distribution shows peak at 109º, which is the conventional value in regular tetrahedron. The width of the peak differs for the equilibrated model vs the melt-quench models, with the latter having less width represents a less distorted tetrahedra for P. In Figure 4(b), the O-Fe-O bond angle shows a broad peak at 95º, ranging from 60º to 180º, representing highly distorted polyhedra for Fe. In this case, there is no discernible difference between the equilibrated and the MQ models. In Figure 4(c), the P-O-Fe bond angle shows a typical bell-shaped distribution seen for bridging bond angle distribution between two polyhedral units sharing a corner, with a peak at ~140º. Here the difference between equilibrated and MQ models can be observed at lower angles from 80º to 110º. Since the MC model is having some percentage of bond angle in range 60-110º, upon equilibration the model still has higher percentage of atoms in range 80-110º. Only upon melt-quenching of MC model, these bond angles are reduced. Hence the magenta curve corresponding to MC_MQ model shows good agreement with the MQ model of Stoch et al. [8]. A similar discussion holds for the polyhedral connection making P-O-P bond angle shown in Figure 4(d).

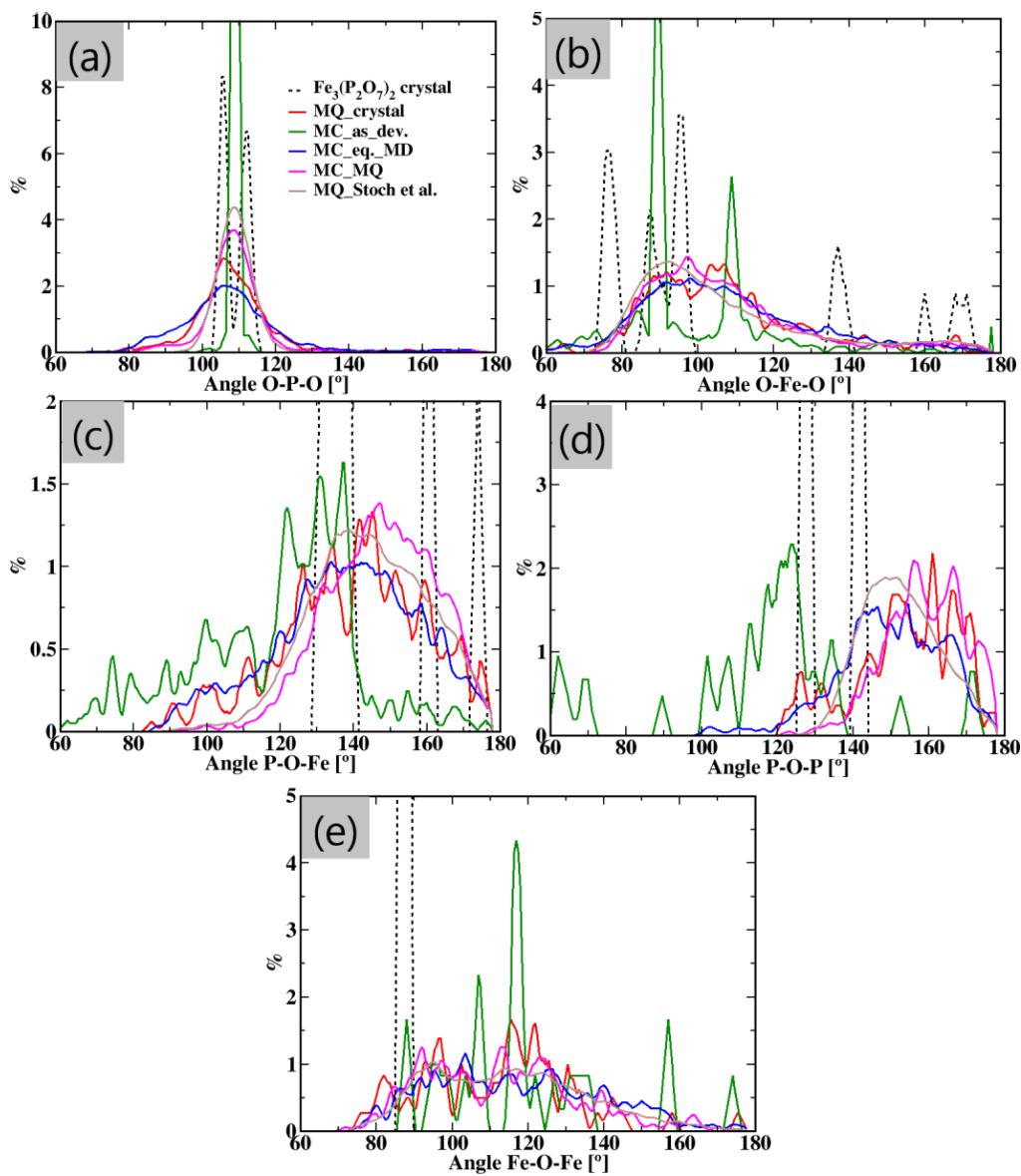

*Figure 4: Bond angle distributions in IPG models*

Finally, the Fe-O-Fe bond angle distribution shown in Figure 4(e) shows a weakly bimodal distribution centred at 95º and 120º. The reason is explained through Figure 5, which shows a snapshot from MC-MQ model showing Fe atom clustering, with two penta-coordinated Fe ($Fe_1$ and $Fe_2$) sharing common polyhedral edge and one Fe ($Fe_3$) in tetrahedral coordination geometry, all sharing a common O vertex (the subscripts of Fe in this paragraph denote labels of Fe as shown in Figure 5). The typical value of Fe-O-Fe bond angle arising from sharing of a polyhedral edge, represented by angle $Fe_1$-O-$Fe_2$ in the snapshot, is ~99º which gives insight about the first of the observed peaks in Fe-O-Fe angle. The other peak at 120º which ranges from 110º to 160º, arises from corner sharing polyhedral units of two Fe's, which comprise of bond angles such as $Fe_1$-O-$Fe_3$ (119.2º) and $Fe_2$-O-$Fe_3$ (140.9º).

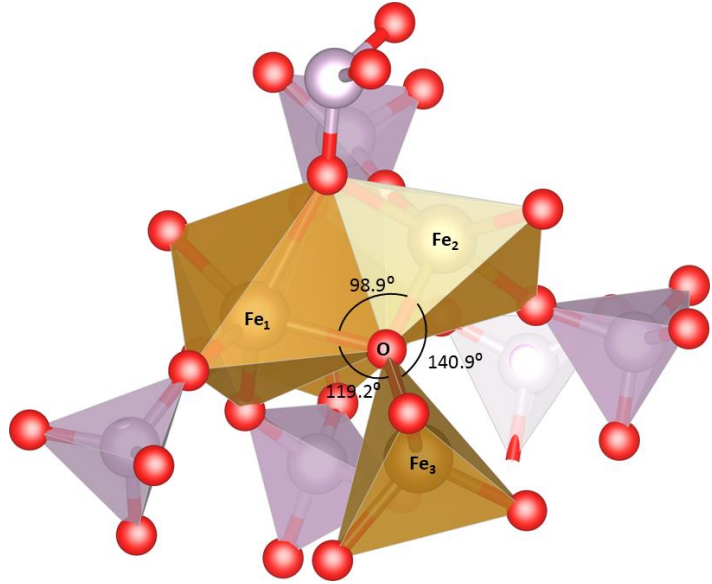

*Figure 5: Snapshot from MC_MQ model showing Fe atom clustering, with two penta-coordinated Fe (Fe$_1$ and Fe$_2$) sharing common polyhedral edge and one Fe (Fe$_3$) in tetrahedral coordination geometry, all sharing a common O vertex. The resultant angles at the common O shows the typical values of Fe-O-Fe bond angles at such vertex. Visualization by VESTA [47]*

In Table 1, the average coordination number for O, P, Fe is estimated for various models considered in this study and compared with models of IPG reported in the literature i.e. the MQ_Stoch et al. model [8], the MC-DFT model [11] and MQ-DFT model [12]. Although the P coordination matches well with experimental value of 3.8 [45] for all the models reported in this study, the Fe CN shows a deviation from 4.7 reported experimentally (4.7 ± 0.3 from ND [45] and 4.74 ± 0.14 from EXAFS [46]). To confirm the departure of MD reported CN for Fe, we also compared the Fe CN from IPG models reported using DFT. The MD reported CN for Fe for the MC_MQ model at 4.3 is below the experimental (4.7) and DFT reported value (4.66 [48] using melt-quench in AIMD, 4.77 [11] using MC+AIMD annealing). It is 4.2 even for the MQ_Stoch et al. model. So, there is a systematic underestimation in CN of Fe reported by MD studies. The reason may lie in the interatomic potential used in MD, addressing which is not in the scope of this study and can be addressed later.

*Table 1: Coordination number comparison of constituents O, P, Fe across different models*

| IPG Models | O CN$_{avg.}$ (ideal glass net. = 2) | P CN$_{avg.}$ (ideal glass net. = 4) | Fe CN$_{avg.}$ |
|---|---|---|---|
| Fe$_3$(P$_2$O$_7$)$_2$ crystal | 2.43 | 4.0 | 6.0 |
| MQ_crystal | 1.99 | 3.87 | 4.08 |
| MC_as_dev. | 1.80 | 3.90 | 3.41 |
| MC_eq_MD | 1.98 | 4.16 | 3.89 |
| MC_MQ | 2.03 | 4.04 | 4.3 |
| MQ_Stoch et al. [8] | 1.95 | 3.89 | 4.2 |
| MC-DFT [11] | 2.08 | 4.0 | 4.77 |
| MC-DFT [12] | 2.10 | 3.96 | 4.62 |
| Expt. [45] | - | $3.8 \pm 0.2$ | $4.7 \pm 0.3$ |

*The superior chemical durability of iron-phosphate glasses*: Table 2 reports the environment around oxygen atom which brings out interesting features of this glass. It is well known that IPG's durability against harsh environment rests on the P-O-Fe bond [49] which is more hydration resistant as compared to the P-O-P bonds that are generally present in glass-formers like $P_2O_5$. Also, while one terminal oxygen per phosphate tetrahedra exist in v-$P_2O_5$, their presence in the network makes it more floppy, when seen using topological constraint theory of glass [50]. So, their percentage should be limited. On the other hand, environments like P-O-$Fe_2$ (CN(O)=3) over-constrain the network and hence their percentages should also be kept in check. Table 2 compares these features in different models of IPG, and in the absence of relevant experimental data MC-DFT model [11] of IPG is used due to parameter-free nature of DFT. The table also reports average oxygen CN of different models in the last column. It can be observed that the MC_MQ model agrees very closely in most of the aspects compared in this table with MC-DFT data from our previous study [11].

The above discussion on Table 2 was purely theoretical in nature. We can also compare the estimates of Table 2 with experimental observations. Experimentally, glass researchers have used EXAFS data to conclude that the concentration of these P-O-Fe links is similar in both crystalline $Fe_3(P_2O_7)_2$ and IPG [51]. The model that agrees best with the crystalline value of 85.72% P-O-Fe bonds is the MC_MQ model with 79.6% of them. Also, the P-O-P bond percentages also agree with each other in these models. Moreover, the experimental observation of P-O-Fe linkages being of type that restricts Fe atom clustering [51] (as in P-O-$Fe_2$) is also clearly highlighted by the MC_MQ model. These arguments are also in line with topological constraint theory of glasses [50] which describes the ideal coordination of oxygen for glass formation as 2. So, linkages of type P-O-$Fe_2$ or P-O-$Fe_3$, do not aide in glass formation. Furthermore, infrared spectrum study of IPG [52] shows that the concentration of P=O (i.e. the terminal oxygen) vanishes in IPG compositions having iron oxide in excess of 33 mol%. The MC_MQ model have the least percentage of these terminal oxygens as also seen in our models developed using DFT [11]. The above discussion highlights the importance of various experimental studies ranging from ND, EXAFS to IR spectra in validating the short-range aspect of the theoretical models.

*Table 2: Bonding distribution around Oxygen in models considered in this study, shedding light on bonding and networking information in these models, the bonds that are typically important for ease of glass formation are labelled f = 0 (representing isostatic case), whereas floppy and rigid*

*network bonding are labelled f < 0 and f > 0, respectively. $O_{NB}$ and $O_B$ stands for non-bridging and bridging oxygen, respectively in the network.*

| Models | f = 0 (ideal) | | f < 0 (floppy) | f > 0 (rigid) | Avg. CN of O |
|---|---|---|---|---|---|
| | P-$O_{NB}$-Fe | P-$O_B$-P | P=$O_{terminal}$ | P-O-$Fe_2$ | |
| | % distribution | | | | |
| $Fe_3(P_2O_7)_2$ crystal | 42.86 | 14.29 | 0 | 42.86 | 2.43 |
| MQ_crystal | 59.6 | 19.64 | 10.5 | 6.25 | 1.99 |
| MC_as_dev. | 57.7 | 7.8 | 25.8 | 4.4 | 1.81 |
| MC_eq_MD | 55.3 | 21.4 | 10.6 | 5.7 | 1.98 |
| MC_MQ | 76.3 | 14.8 | 2.8 | 3.3 | 2.02 |
| MQ_Stoch et al. [8] | 54.5 | 22.1 | 10.8 | 3.6 | 1.95 |
| MC-DFT [11] | 74 | 15 | 2 | 7 | 2.08 (HSE) |

## B. Medium-range properties
### 1. Structure factor analysis

The medium range structure of IPG can be scrutinized with the help of structure factor, S(q). Figure 6 shows the calculated neutron structure factor of different IPG models considered in this study and the experimental S(q) from [45]. The discussion on structure factor can be divided into two parts. First, the FSDP is observed in the region from 1 to 2.5 Å$^{-1}$. The magnitude of the experimental S(q) and the calculated S(q)s may not be comparable because it is not possible to account for all the contributions to the former, but the magnitude of calculated S(q) can be compared amongst IPG models as we compare the same contributions. The features of the calculated S(q) of IPG models like FSDP, shoulder in the second peak and the second peak itself, the third peak etc. develop around corresponding peaks in the S(q) of the crystal. The failure of MQ_crystal model in correctly modelling the MR structure is clearly visible in FSDP range. Also, the MC_as_dev. model fails to correctly reproduce the FSDP, which implies the need for equilibration of the potential-free MC model. The MC_eq_MD model, MC_MQ model and the MQ_Stoch et al. model, all correctly reproduce the peak position of FSDP at ~1.75 Å$^{-1}$ although the differing magnitude of the peak may point to how much a model is ordered/disordered in the MR range. More about the shape of FSDP and its origin from rings viewpoint will be discussed later.

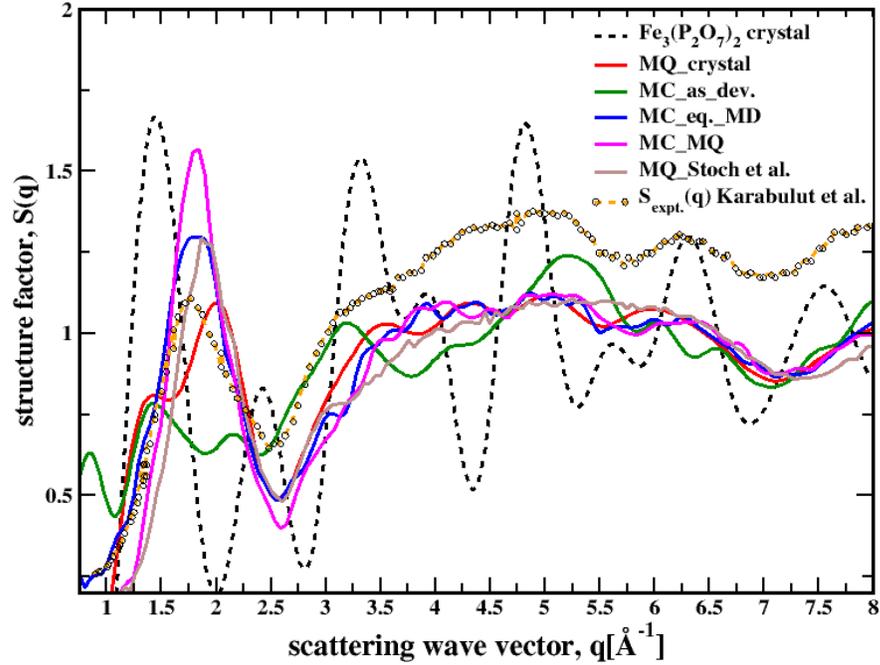

*Figure 6: Neutron structure factor of various models considered in this study and a comparison with experimental data from* [45]

## 2. Void size distribution

The void size distribution of a glass can help in characterising its medium-range structure. Figure 7 (a) shows a schematic of the empty space that is referred here. The code used for void estimation is given at [53]. In Figure 7(b), the void size distribution of the models considered in this study is reported. The void fraction which is the ratio of space occupied by voids to the total space available inside the simulation box, is also reported alongside the each of the graphs which are stacked vertically for better comparison. A homogeneously mixed glass structure like that of IPG is expected to have a bell-shaped distribution of void sizes having a most probable void size that occurs roughly at the centre of the distribution. Also, crystalline structure is expected to have discrete peaks in void distribution due to its periodicity in its packing. The graphs in Figure 7 (b) clearly highlights these features. The MC_as_dev. structure shows a larger variation of void sizes which upon equilibration (or melt-quench) improves its packing as also shown by a decreased void-fraction. In terms of void-fraction and void size distribution, the MC_eq._MD model and MC_MQ model have identical results.

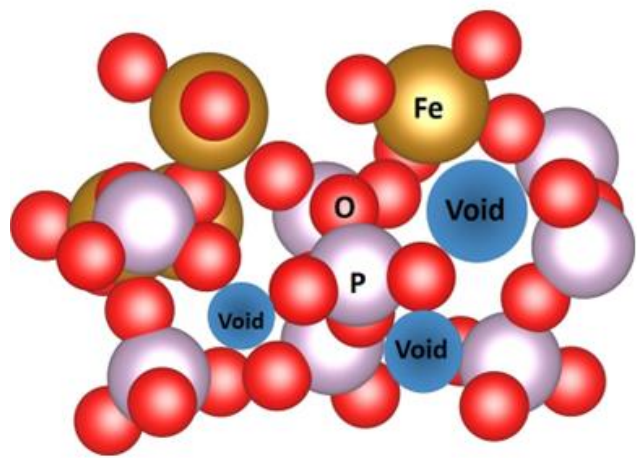
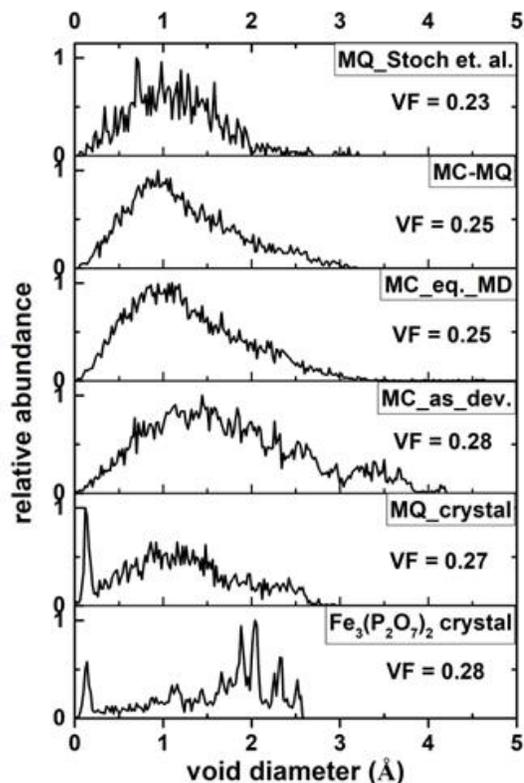

*Figure 7: (a) Schematic of the empty space inside glass, here spherical voids are used to fill the empty space, (b) void size distribution of the models considered in this study*

### 3. Detailed rings analysis

Next, the rings size distribution of IPG models calculated using R.I.N.G.S. code [54] is shown in Figure 8. The rings definition used in the calculations is after Guttman [55]. According to it, ring is defined as the shortest path traversed along the bonds and its size is the number of atoms (also called nodes) contained in the closed path.

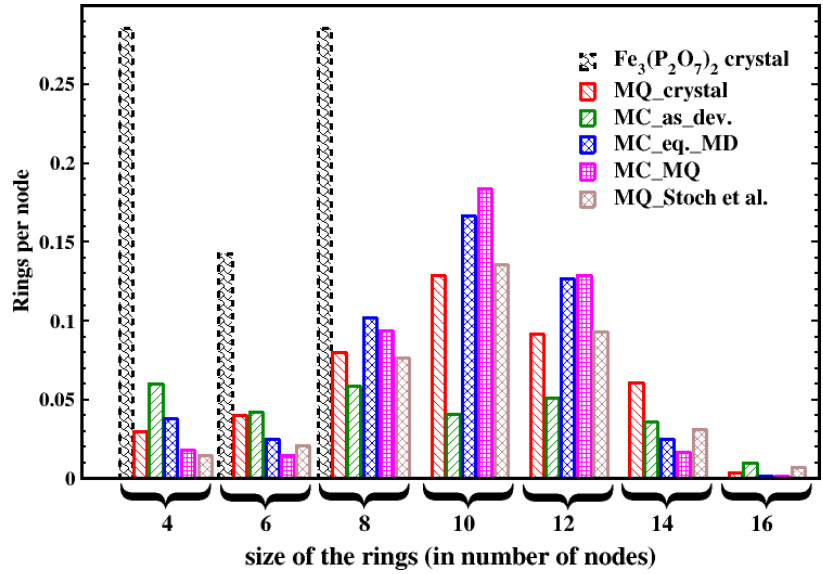

*Figure 8: Rings size distribution in various IPG models considered in this study*

The rings size distribution of the $Fe_3(P_2O_7)_2$ crystal model shows only three possibilities of rings. For easier understanding the crystal structure (Figure 9(a)) and the rings encountered in this system (Figure 9 (b,c)) are shown in Figure 9. The face-sharing Fe's in the crystal structure results in 4-node rings in this system. Moreover, only 4, 6 and 8-node rings are present although the ring counts of each of these rings is very high as compared to the glasses. On the other hand, the distribution in IPG models shows possibilities from 4 to 16 node rings. Hence, the topological complexity (in terms of varieties of ring sizes) of the IPG models is far more than crystalline system as also known for other glass systems like silica [34,56] etc. As we have focused on …A-B-A-C… type of rings, only even membered rings are possible. The distribution for IPG is found to peak at 10-node ring size around which it decays on either side. Also, the absence of FSDP in MC_as_dev. model and MQ_crystal model can be correlated with rings distribution. MC_as_dev. model shows no peak in the rings distribution and fails to produce any FSDP in S(q). The MQ_crystal model though has a peak in rings distribution at 10-node size yet the intensities of smaller sized rings (4, 6) are higher as compared to MC_eq_MD, MC_MQ and MQ_Stoch et al. models. The higher intensity of smaller rings, particularly the 4-node rings over-strains (f > 0 type) the glass network resulting in higher undesirable P-O-$Fe_2$ type of configurations, which are the edge-shared or face-shared configurations of Fe polyhedral units. These rings are in higher count even in MC_eq_MD model, showing the tendency of Fe atoms to cluster in the network. Only upon melt-quenching, these clusters are reduced as shown in the reduced ring count in MC_MQ and MQ_Stoch et al models. Among these two models, the 10-node rings for MC_MQ models are higher than MQ_Stoch et al. model, possibly showing more order and hence the higher magnitude of FSDP for MC_MQ model.

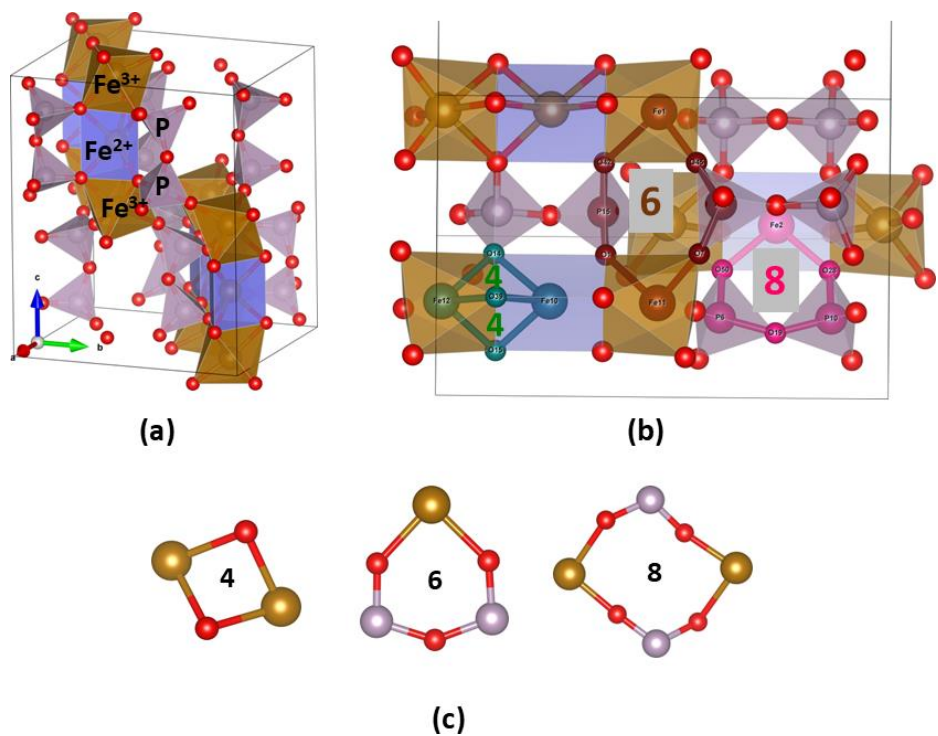

*Figure 9: (a) $Fe_3(P_2O_7)_2$ crystal structure with face sharing type $Fe^{3+}$-$Fe^{2+}$ connections, $Fe^{2+}$ trigonal prismatic geometry coloured in blue and $Fe^{3+}$ octahedral geometry coloured in golden brown, purple tetrahedral geometry is of pentavalent P (b) 4, 6, 8-node ring sizes present in the crystal shown with green, maroon and pink bonding respectively, the involved atoms are labelled and (c) 4, 6 and 8-node rings shown separately, Fe: golden brown, O: red, P: purple. Visualization by VESTA* [47]

    The topological complexity of one of the models, i.e. the MC_MQ model is shown in Figure 10, where different representative configurations of the rings for each ring size are shown. In general, for a ring of size n-node, $n-2$ distinct configurations are encountered in the models, except for 4-node rings where only one configuration is found. Of course, to find all these configurations the model should be large enough. Also, for larger ring sizes such as 12 and 14 the rings are puckered in nature, which is not visually obvious from the 2D images shown in Figure 10. For 10-node ring size the number of rings in each of the configurations is also counted and enumerated in Table 3 for selected models. The table highlights the fact that long chains of either only P or only Fe atoms are less probable in the models, as seen by lower counts in 5P and 5Fe rings. All the models show the highest ring count for 3P-2Fe types of rings followed by its complementary 2P-3Fe type of rings. This implies Fe homogeneously mixes throughout the glass network. This inference is an important one and can be used to qualitatively explain the low electronic conductivity of IPG [57]. Electronic conductivity of IPG is explained through polaronic transport mechanism [58–61]. Polaronic transport critically depends on the polaron hopping length and the concentration of Fe. Large polaron hopping lengths facilitates polaronic conductivity, which requires continuous chains/rings of type …Fe-O-Fe-O-Fe-O… so that polaron hopping can take place from $Fe^{2+}$ and $Fe^{3+}$ over large distances. But these rings are found in very low concentration in all the models as seen from the last two columns of Table 3. So, the homogenous

mixing of iron in the phosphate network contributes towards small hopping lengths and hence low polaronic conductivity of IPG. Also, the MQ_Stoch et al. model shows 8 rings of 5Fe type as well, which were not seen in other models. This may be because the model is very large as compared to other models in this study, which might help in accommodating all possible configurations, even the less energetically favoured ones in some counts. So, the topological analysis as shown in Table 3 and Figure 10 is useful not only in modelling MRO in IPG but also in explaining experimentally observed results from atomistic viewpoint.

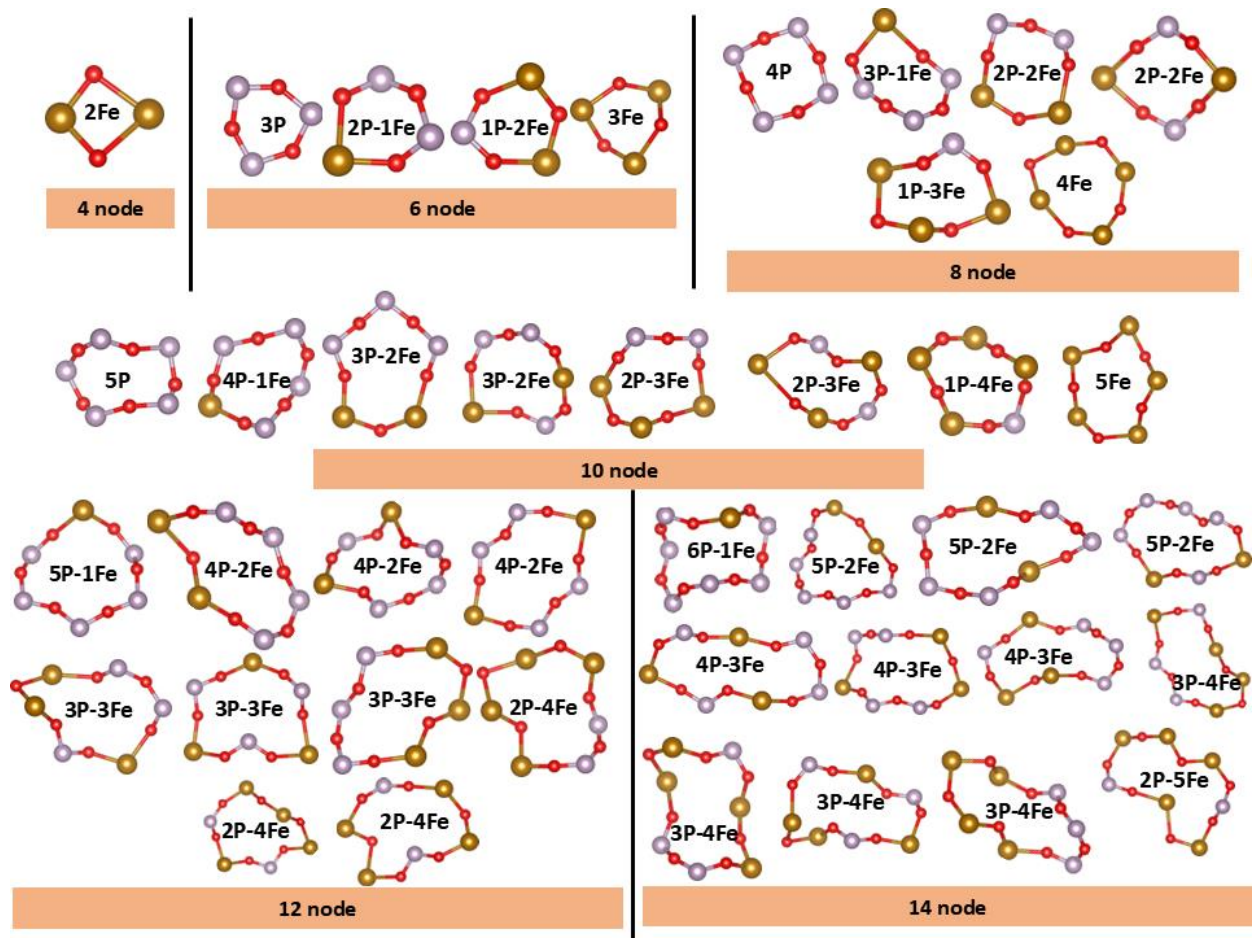

*Figure 10: Different representative configurations of rings found for different ring sizes in the MC_MQ IPG model. The number of P and Fe atoms involved is written for each configuration. For configurations with same P-Fe atoms all distinct arrangements of atoms are shown. (Fe: golden brown, P: purple, O: red) Visualization by VESTA* [47]

*Table 3: Table showing the number of rings in each of the different types of configurations present in 10-node ring size in IPG models (phosphorous: purple, oxygen: red, iron: golden brown)*

| | 5P | 4P-1Fe | 3P-2Fe | | 2P-3Fe | | 1P-4Fe | 5Fe |
|---|---|---|---|---|---|---|---|---|
| Ring type (P-Fe atoms) | 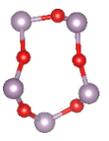 | 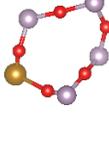 | 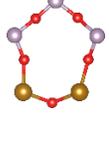 | 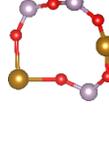 | 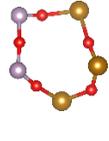 | 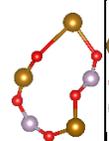 | 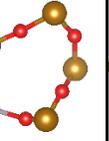 | 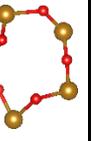 |
| MQ_crystal | 0 | 8 | 44 | | 32 | | 3 | 0 |
| MC_eq_MD | 0 | 58 | 523 | | 227 | | 10 | 0 |
| MC_MQ | 2 | 52 | 563 | | 277 | | 7 | 0 |
| MQ_Stoch et al. | 42 | 744 | 3313 | | 2836 | | 409 | 8 |

The individual ring size structure factor, $S_n(q)$ is calculated using R.I.N.G.S. [54] by only considering those atoms as scattering centres which are contributing towards rings formation of a particular size. Hence $S_4(q)$, $S_6(q)$, $S_8(q)$, $S_{10}(q)$, $S_{12}(q)$, $S_{14}(q)$ and $S_{16}(q)$ (subscripts denote ring size) are calculated for IPG models: MC_eq_MD, MC_MQ, MQ_crystal and MQ_Stoch et al.. The $S_n(q)$'s brings out an interesting picture of formation of FSDP in S(q). For each of the ring sizes, the FSDP falls in the same range (1, 2.5 Å) as the FSDP of total S(q) and the peak positions follow the trend: $S_4(q) > S_6(q) > S_8(q) > S_{10}(q) \approx S_{total}(q) > S_{12}(q) > S_{14}(q) > S_{16}(q)$. Correlating with rings size distribution trend, the FSDP in total S(q) occurs at the same position as the FSDP of 10-node rings. As the $q_{peak}$ tells about the size of maximally repeating cluster size in the models, the coincidence of the two peaks implies the rings distribution can give insight about the construction of the FSDP in total S(q). Moreover, FSDP's of remaining rings sizes occur according to their sizes, i.e. for rings larger than 10-node (12, 14) the peak position falls on the smaller q side and for smaller ring sizes (4, 6, 8) the peak position falls on larger q side. This interesting correlation has also been observed for silica-based glasses recently [17].

In this study, we do inverse fitting of the individual $S_n(q)$'s so that their contribution in construction of total S(q) can be estimated by their respective weight-coefficients assuming the following relation:

$$S_{exp}(q) = A_1 * S_4(q) + A_2 * S_6(q) + A_3 * S_8(q) + A_4 * S_{10}(q) + A_5 * S_{12}(q) + A_6 * S_{14}(q)$$

where the $A_i$'s are unknown weight-coefficients and $S_n(q)$'s are known from calculations on theoretical model and $S_{exp}(q)$ is obtained from neutron diffraction experiments from [45]. All possible ring sizes (except 16-node, due to very small contribution) are considered in this fitting. The RHS of above equation is the calculated $S_{calc}(q)$, and the fitting is done by minimizing the mean squared error from $S_{exp}(q)$. From Figure 11, it can be seen that the intensities of $S_n(q)$ peaks for our models is less than that of the large model developed by Pawel Stoch et al. [8]. Also, the $S_n(q)$'s shoots up at small q values relatively earlier in our models which may also be due to the small sizes of our models. This leads to convergence problems in fitting if full range of FSDP [1, 2.5] is taken. So, in the fitting analysis presented in Figure 12, for (a) MC_MQ model and (b) MQ_Stoch et al. model, this q-range is ignored and fitting is started from 1.5 and 1.3 Å$^{-1}$ respectively. The rings distribution of these models obtained through R.I.N.G.S. [54] and $A_i$'s obtained from fitting are plotted in a bar chart in Figure 12 (c) and (d). The $A_i$'s obtained from fitting qualitatively agrees with the bell-shaped rings distribution shown by these models. But there is quantitative difference seen for 8-node, 12-node and 14-node rings. This may sound as a limitation of the current models as there is no way to control the number of these rings through simulation at present. However, the role of rings distribution as a structural descriptor for medium-range order is established and its connection to the experimental structure-factor can allow for validation of the theoretical models at the medium-range length-scale.

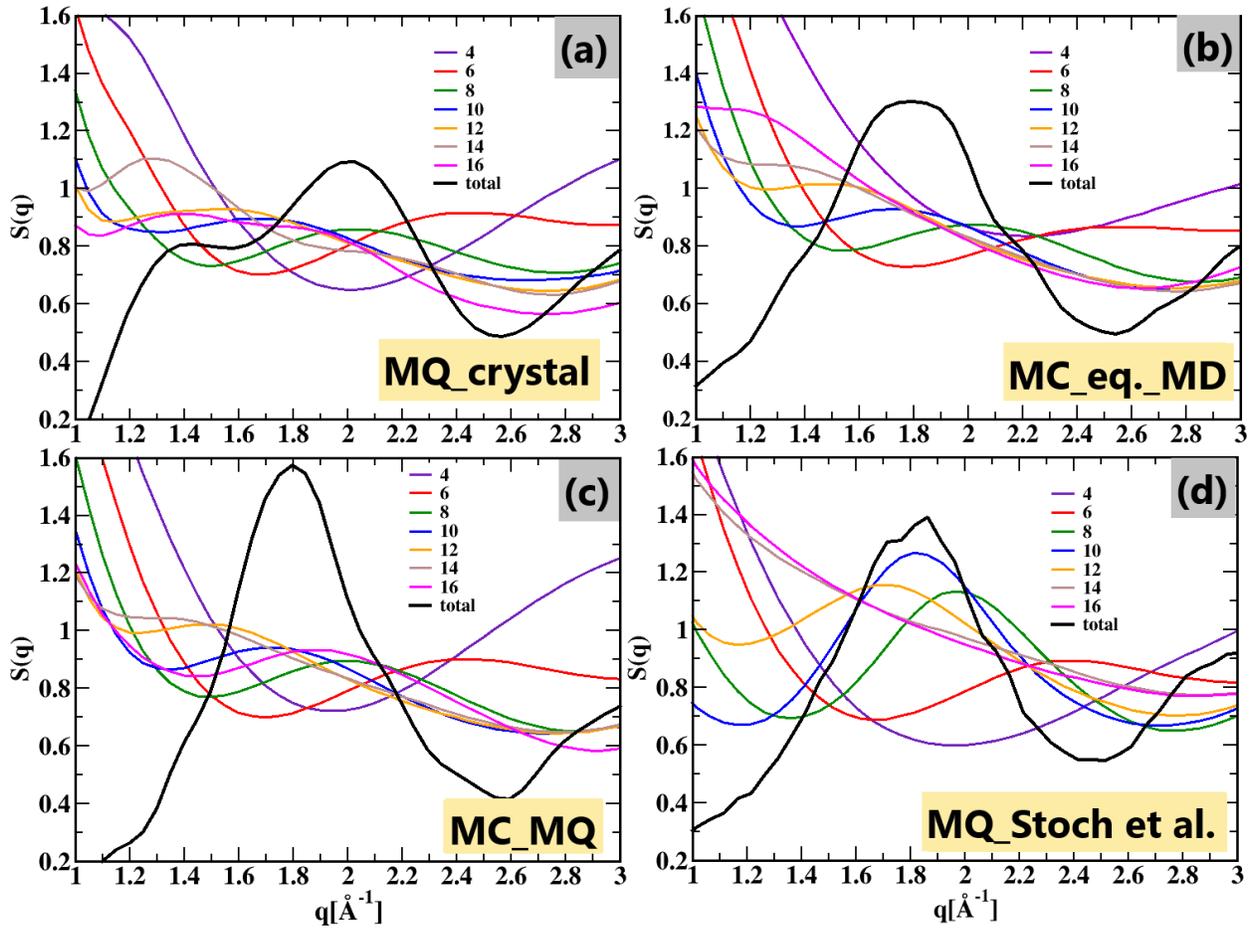

*Figure 11: A comparison of individual ring size structure factors $S_n(q)$'s for the glass models considered in this study, curve shown in black is the total neutron structure factor S(q) for the model. The graphs show only the FSDP range of q values i.e. [1, 3] Å$^{-1}$*

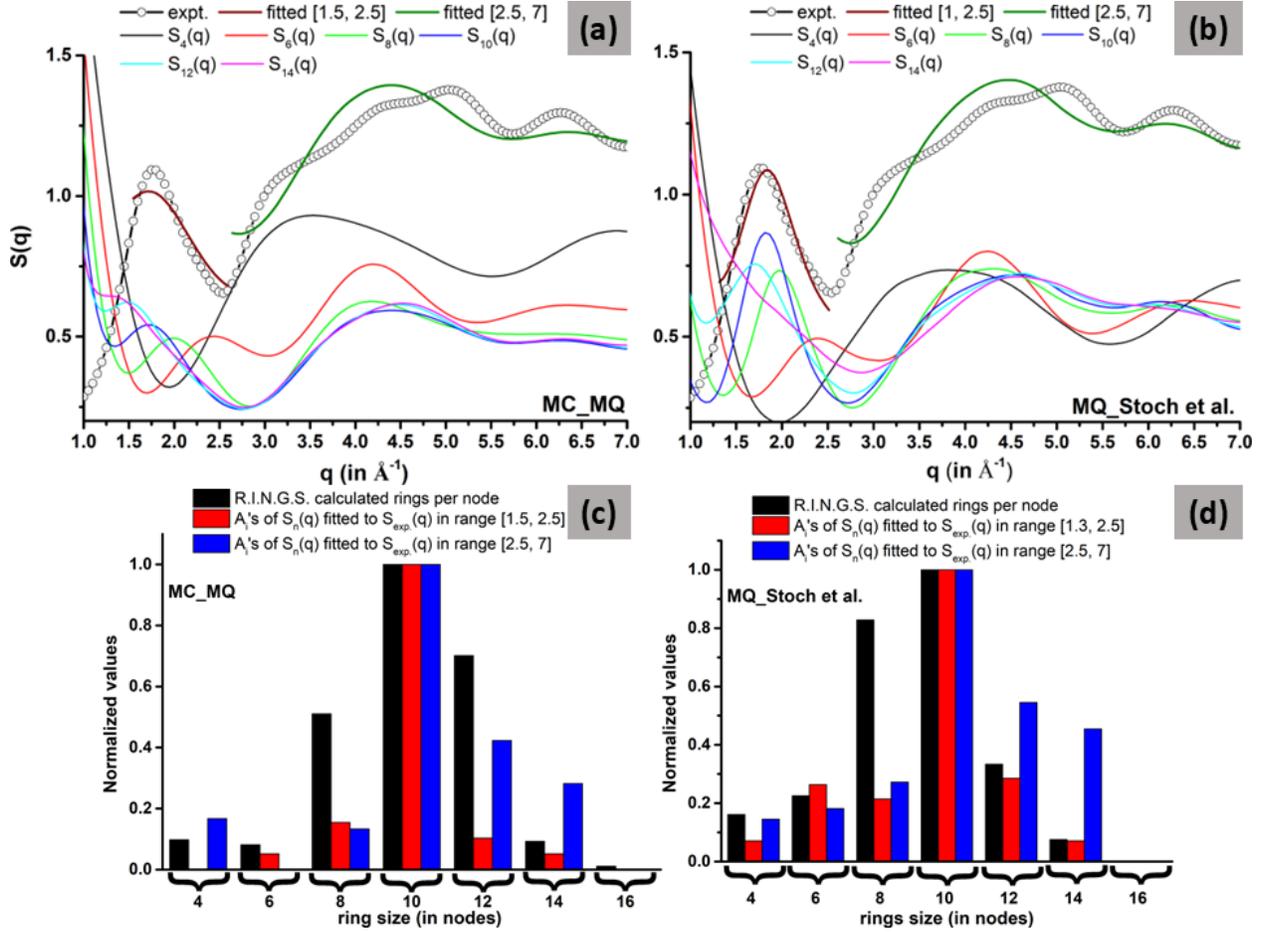

*Figure 12: Fitting of $S_n(q)$'s to $S_{exp}(q)$ for (a) MC_MQ model and (b) MQ_Stoch et al. model, corresponding rings distributions and $A_i$'s for (c) MC_MQ and (d) MQ_Stoch et al. models*

# IV. Conclusion

The study explores IPG models obtained through various simulation procedures and from different initial configurations, for a better understanding of the glass structure, particularly in the medium range. The comparison show that melt-quench model obtained using the MC_as_dev model as initial randomized configuration shows better overall agreement with the experimental data. The comparison also highlights the systematic deviation of all MD processed models from experimental data as well as DFT calculated values, when coordination number of Fe is concerned. This highlights issue with interatomic potential for IPG. Also, the comparison of all the structural properties with that of crystalline iron phosphate also shows the resemblance of IPG with the crystalline system considered. Structural properties of MQ_crystal model show some crystalline signatures left even after melt-quenching which corroborates the existing understanding that choosing crystals as initial configurations may not be suitable for homogeneous glassy models.

The topological analysis of the rings present in the IPG models gives insights on the homogeneous mixing of the Fe in the phosphate network, which is qualitatively shown to explain the low polaronic conductivity of IPG. The calculation of structure factor corresponding to different size of the rings not only shines light on the make-up of FSDP in total S(q), but also helps in giving the rings distribution a physical motivation through experimental structure factor data. Hence it proves to be a much-needed descriptor for MRO in complex glasses as well.

The rings data presented in the study also validated the hybrid method of developing complex glass models using MC+CMD, by satisfactorily reproducing the medium range structure as well. The MC+AIMD(DFT) way shown in [11], produced structures that lacked a good estimation of structure factor and rings distribution owing to their smaller size. Here it is shown that larger models can indeed reproduce the MRO as seen using the FSDP and rings distribution, of course at the cost of electronic level accuracy that is available in DFT. The subsequent rings configuration analysis enhances our understanding of MRO in this important glass.

**References**

[1]   A. C. Wright and M. F. Thorpe, Eighty years of random networks, Phys. Status Solidi Basic Res. **250**, 931 (2013).

[2]   A. K. Varshneya and J. C. Mauro, *Fundamentals of Inorganic Glasses* (Elsevier, 2019).

[3]   B. Al-Hasni and G. Mountjoy, Structural investigation of iron phosphate glasses using molecular dynamics simulation, J. Non. Cryst. Solids **357**, 2775 (2011).

[4]   K. Joseph, K. Jolley, and R. Smith, Iron phosphate glasses: Structure determination and displacement energy thresholds, using a fixed charge potential model, J. Non. Cryst. Solids **411**, 137 (2015).

[5]   K. Jolley and R. Smith, Iron phosphate glasses: Structure determination and radiation tolerance, Nucl. Instruments Methods Phys. Res. Sect. B Beam Interact. with Mater. Atoms **374**, 8 (2016).

[6]   K. Jolley and R. Smith, Radiation tolerance of iron phosphate: A study of amorphous and crystalline structures, J. Nucl. Mater. **479**, 347 (2016).

[7]   K. Joseph, M. C. Stennett, N. C. Hyatt, R. Asuvathraman, C. L. Dube, A. S. Gandy, K. V. Govindan Kutty, K. Jolley, P. R. Vasudeva Rao, and R. Smith, Iron phosphate glasses: Bulk properties and atomic scale structure, J. Nucl. Mater. **494**, 342 (2017).

[8]   P. Goj and P. Stoch, Molecular dynamics simulations of $P_2O_5$-$Fe_2O_3$-FeO-$Na_2O$ glasses, J. Non. Cryst. Solids **500**, 70 (2018).

[9]   P. Stoch, W. Szczerba, W. Bodnar, M. Ciecinska, A. Stoch, and E. Burkel, Structural properties of iron-phosphate glasses: Spectroscopic studies and ab initio simulations, Phys. Chem. Chem. Phys. **16**, 19917 (2014).

[10]  P. Stoch, A. Stoch, M. Ciecinska, I. Krakowiak, and M. Sitarz, Structure of phosphate and iron-phosphate glasses by DFT calculations and FTIR/Raman spectroscopy, J. Non. Cryst. Solids **450**, 48 (2016).

[11] S. Singh, M. Dholakia, and S. Chandra, Toward Atomistic Understanding of Iron Phosphate Glass: A First-Principles-Based Density Functional Theory Modeling and Study of Its Physical Properties, J. Phys. Chem. B **128**, 3258 (2024).

[12] S. S. Jena, S. Chandra, S. Singh, and G. Kaur, Characterization of Iron Phosphate glass from melt-quench simulations using ab-initio molecular dynamics, J. Non. Cryst. Solids **637**, 123035 (2024).

[13] D. Dahal, H. Warren, and P. Biswas, On the Origin and Structure of the First Sharp Diffraction Peak of Amorphous Silicon, Phys. Status Solidi Basic Res. **258**, 1 (2021).

[14] C. Crupi, G. Carini, G. Ruello, and G. D'Angelo, Intermediate range order in alkaline borate glasses, Philos. Mag. **96**, 788 (2016).

[15] J. H. Lee and S. R. Elliott, Simulations of void-filled vitreous silica to interpret the origin of the first sharp diffraction peak, Phys. Rev. B **50**, 5981 (1994).

[16] S. R. Elliott, Extended-range order, interstitial voids and the first sharp diffraction peak of network glasses, J. Non. Cryst. Solids **182**, 40 (1995).

[17] Q. Zhou, Y. Shi, B. Deng, J. Neuefeind, and M. Bauchy, Experimental method to quantify the ring size distribution in silicate glasses and simulation validation thereof, Sci. Adv. **7**, 1 (2021).

[18] Q. Zhou, Y. Shi, B. Deng, T. Du, L. Guo, M. M. Smedskjaer, and M. Bauchy, Revealing the medium-range structure of glassy silica using force-enhanced atomic refinement, J. Non. Cryst. Solids **573**, 121138 (2021).

[19] W. H. Zachariasen, The atomic arrangement in glass, J. Am. Chem. Soc. **54**, 3841 (1932).

[20] P. Y. Huang et al., Direct Imaging of a Two-Dimensional Silica Glass on Graphene, Nano Lett. **12**, 1081 (2012).

[21] Y. Shi, J. Neuefeind, D. Ma, K. Page, L. A. Lamberson, N. J. Smith, A. Tandia, and A. P. Song, Ring size distribution in silicate glasses revealed by neutron scattering first sharp diffraction peak analysis, J. Non. Cryst. Solids **516**, 71 (2019).

[22] A. F. Firooz, R. Christensen, C. A. N. Biscio, and M. M. Smedskjaer, Characterizing medium-range order structure of binary silicate glasses using ring analysis and persistent homology, J. Am. Ceram. Soc. 1 (2024).

[23] P. S. Salmon, A. Zeidler, M. Shiga, Y. Onodera, and S. Kohara, Ring compaction as a mechanism of densification in amorphous silica, Phys. Rev. B **107**, 1 (2023).

[24] M. Shiga, A. Hirata, Y. Onodera, and H. Masai, Ring-originated anisotropy of local structural ordering in amorphous and crystalline silicon dioxide, Commun. Mater. **4**, 91 (2023).

[25] N. M. A. Krishnan, B. Wang, Y. Le Pape, G. Sant, and M. Bauchy, Irradiation- vs. vitrification-induced disordering: The case of α -quartz and glassy silica, J. Chem. Phys. **146**, 204502 (2017).

[26] Y. Yang, H. Tokunaga, K. Hayashi, M. Ono, and J. C. Mauro, Understanding thermal

expansion of pressurized silica glass using topological pruning of ring structures, J. Am. Ceram. Soc. **104**, 114 (2021).

[27] X. Bidault, S. Chaussedent, W. Blanc, and D. R. Neuville, Deformation of silica glass studied by molecular dynamics: Structural origin of the anisotropy and non-Newtonian behavior, J. Non. Cryst. Solids **433**, 38 (2016).

[28] F. Font-Clos, M. Zanchi, S. Hiemer, S. Bonfanti, R. Guerra, M. Zaiser, and S. Zapperi, Predicting the failure of two-dimensional silica glasses, Nat. Commun. **13**, 1 (2022).

[29] D. Ormrod Morley and M. Wilson, Controlling disorder in two-dimensional networks., J. Phys. Condens. Matter **30**, 50LT02 (2018).

[30] M. Wilson, The effects of topology on the structural, dynamic and mechanical properties of network-forming materials, J. Phys. Condens. Matter **24**, (2012).

[31] A. Nakano, R. Kalia, and P. Vashishta, First sharp diffraction peak and intermediate-range order in amorphous silica: finite size effects in molecular dynamics simulations, J. Non. Cryst. Solids **171**, 157 (1994).

[32] M. Ijjaali, G. Venturini, R. Gerardin, B. Malaman, and C. Gleitzer, Synthesis, Structure and Physical Properties of a Mixed-Valence Iron Diphosphate Fe3(P2O7)2: First Example of Trigonal Prismatic Fe2+ with O2-Ligands, Eur. J. Solid State Inorg. Chem. **28**, 983 (1991).

[33] A. Jain et al., Commentary: The materials project: A materials genome approach to accelerating materials innovation, APL Mater. **1**, (2013).

[34] S. Singh and S. Chandra, Developing atomistic glass models using potential-free Monte Carlo method: From simple to complex structures, Comput. Mater. Sci. **202**, 110943 (2022).

[35] S. Plimpton, P. Crozier, and A. Thompson, *LAMMPS-Large-Scale Atomic/Molecular Massively Parallel Simulator*.

[36] R. A. Buckingham, The classical equation of state of gaseous helium, neon and argon, Proc. R. Soc. London. Ser. A. Math. Phys. Sci. **168**, 264 (1938).

[37] J. E. Lennard-Jones, Cohesion, Proc. Phys. Soc. **43**, 461 (1931).

[38] J. F. Ziegler and J. P. Biersack, *The Stopping and Range of Ions in Matter*, in *Treatise on Heavy-Ion Science* (Springer US, Boston, MA, 1985), pp. 93–129.

[39] F. H. Stillinger and T. A. Weber, Computer simulation of local order in condensed phases of silicon, Phys. Rev. B **31**, 5262 (1985).

[40] L. Verlet, Computer "Experiments" on Classical Fluids. I. Thermodynamical Properties of Lennard-Jones Molecules, Phys. Rev. **159**, 98 (1967).

[41] L. Verlet, Computer "experiments" on classical fluids. II. Equilibrium correlation functions, Phys. Rev. **165**, 201 (1968).

[42] W. C. Swope, H. C. Andersen, P. H. Berens, and K. R. Wilson, A computer simulation method for the calculation of equilibrium constants for the formation of physical clusters of molecules: Application to small water clusters, J. Chem. Phys. **76**, 637 (1982).

[43] L. Deng and J. Du, Development of effective empirical potentials for molecular dynamics simulations of the structures and properties of boroaluminosilicate glasses, J. Non. Cryst. Solids **453**, 177 (2016).

[44] D. Frenkel and B. Smit, *Understanding Molecular Simulation From Algorithms to Applications* (Academic Press Inc., 2002).

[45] M. Karabulut, G. K. Marasinghe, C. S. Ray, G. D. Waddill, D. E. Day, Y. S. Badyal, M. L. Saboungi, S. Shastri, and D. Haeffner, A high energy x-ray and neutron scattering study of iron phosphate glasses containing uranium, J. Appl. Phys. **87**, 2185 (2000).

[46] D. P. Dutta, M. Roy, R. K. Mishra, S. S. Meena, A. Yadav, C. P. Kaushik, and A. K. Tyagi, Structural investigations on Mo, Cs and Ba ions-loaded iron phosphate glass for nuclear waste storage application, J. Alloys Compd. **850**, 156715 (2021).

[47] K. Momma and F. Izumi, VESTA: A three-dimensional visualization system for electronic and structural analysis, J. Appl. Crystallogr. **41**, 653 (2008).

[48] S. S. Jena, S. Singh, and S. Chandra, *Developing Atomistic Models of Iron-Phosphate Glass Using Ab-Initio Molecular Dynamics Simulations*, in *International Conference on Advances in Glass & Glass-Ceramics (ICAGGC 2022)* (2022), p. 28.

[49] X. Yu, D. E. Day, G. J. Long, and R. K. Brow, Properties and structure of sodium - Iron phosphate glasses, J. Non. Cryst. Solids **215**, 21 (1997).

[50] J. C. Mauro, Topological constraint theory of glass, Am. Ceram. Soc. Bull. **90**, 31 (2011).

[51] C. H. Booth, P. G. Allen, J. J. Bucher, N. M. Edelstein, D. K. Shuh, G. K. Marasinghe, M. Karabulut, C. S. Ray, and D. E. Day, Oxygen and phosphorus coordination around iron in crystalline ferric ferrous pyrophosphate and iron-phosphate glasses with $UO_2$ or $Na_2O$, J. Mater. Res. **14**, 2628 (1999).

[52] G. Wang, Y. Wang, and B. Jin, *Structural Properties of Sodium-Iron Phosphate Glasses*, in *Proc. SPIE 2287, Properties and Characteristics of Optical Glass III*, Vol. 2287 (1994), pp. 214–225.

[53] M. Usdin, *MOF Explorer at Https://Mausdin.Github.Io/MOFsite/MofPage.Html*, https://mausdin.github.io/MOFsite/mofPage.html.

[54] S. Le Roux and P. Jund, Ring statistics analysis of topological networks: New approach and application to amorphous $GeS_2$ and $SiO_2$ systems, Comput. Mater. Sci. **49**, 70 (2010).

[55] L. Guttman, Ring structure of the crystalline and amorphous forms of silicon dioxide, J. Non. Cryst. Solids **116**, 145 (1990).

[56] S. S. Jena, S. Singh, and S. Chandra, Characterizing MRO in atomistic models of vitreous $SiO2$ generated using ab-initio molecular dynamics, Appl. Phys. A **129**, 742 (2023).

[57] A. Bafti, S. Kubuki, H. Ertap, M. Yüksek, M. Karabulut, A. Moguš-Milanković, and L. Pavić, Electrical Transport in Iron Phosphate-Based Glass-(Ceramics): Insights into the Role of $B2O3$ and $HfO2$ from Model-Free Scaling Procedures, Nanomaterials **12**, 639 (2022).

[58] B. Šantić, A. Moguš-Milanković, and D. E. Day, The dc electrical conductivity of iron phosphate glasses, J. Non. Cryst. Solids **296**, 65 (2001).

[59] A. Šantić and A. Moguš-Milanković, Charge carrier dynamics in materials with disordered structures: A case study of iron phosphate glasses, Croat. Chem. Acta **85**, 245 (2012).

[60] A. Moguš-Milanković, A. Šantić, A. Gajović, and D. E. Day, Electrical properties of sodium phosphate glasses containing Al2O3 and/or Fe2O3. Part II, J. Non. Cryst. Solids **296**, 57 (2001).

[61] Y. M. Moustafa, K. El-Egili, H. Doweidar, and I. Abbas, Structure and electric conduction of Fe2O3-P 2O5 glasses, Phys. B Condens. Matter **353**, 82 (2004).